%% file: paper.tex
\documentclass[sigconf]{aamas}

\usepackage{balance} % for balancing columns on the final page
\usepackage[linesnumbered,ruled,vlined]{algorithm2e}
\usepackage{caption}
\usepackage{subcaption}
\usepackage{bbm}
\usepackage{tikz}
\usetikzlibrary{shapes, arrows, positioning}
\usepackage{graphicx}
\usepackage{soul}
\usepackage[normalem]{ulem}
\usepackage{hyperref}

\usepackage{amsmath}
\usepackage{amsthm}

\theoremstyle{definition}

\newcommand{\Z}{\mathbb{Z}}
\newcommand{\bd}[1]{\mathbf{#1}}
\newcommand{\bth}{\boldsymbol{\theta}}

\newcommand{\Mod}[1]{\ (\mathrm{mod}\ #1)}

\def\equationautorefname~#1\null{Equation~(#1)\null}

\setcopyright{ifaamas}
\acmConference[AAMAS '24]{Proc.\@ of the 23rd International Conference
on Autonomous Agents and Multiagent Systems (AAMAS 2024)}{May 6 -- 10, 2024}
{Auckland, New Zealand}{N.~Alechina, V.~Dignum, M.~Dastani, J.S.~Sichman (eds.)}
\copyrightyear{2024}
\acmYear{2024}
\acmDOI{}
\acmPrice{}
\acmISBN{}

\acmSubmissionID{70}

\title[Privacy-preserving ABM]{Private Agent-Based Modeling}

\author{Ayush Chopra}
\affiliation{
  \institution{Massachusetts Institute of Technology}
  \city{Cambridge}
   \country{USA}}
\email{ayushc@mit.edu}

\author{Arnau Quera-Bofarull}
\affiliation{
  \institution{University of Oxford}
  \city{Oxford}
  \country{UK}}
\email{arnau.quera-bofarull@cs.ox.ac.uk}

\author{Nurullah Giray-Kuru}
\affiliation{
  \institution{Massachusetts Institute of Technology}
  \city{Cambridge}
  \country{USA}}
\email{ngkuru@mit.edu}

\author{Michael Wooldridge}
\affiliation{
  \institution{University of Oxford}
  \city{Oxford}
  \country{UK}}
\email{michael.wooldridge@cs.ox.ac.uk}

\author{Ramesh Raskar}
\affiliation{
  \institution{Massachusetts Institute of Technology}
  \city{Cambridge}
  \country{USA}}
\email{raskar@media.mit.edu}

\begin{abstract}

The practical utility of agent-based models in decision-making relies on their capacity to accurately replicate populations while seamlessly integrating real-world data streams. Yet, the incorporation of such data poses significant challenges due to privacy concerns. To address this issue, we introduce a paradigm for private agent-based modeling wherein the simulation, calibration, and analysis of agent-based models can be achieved without centralizing the agents' attributes or interactions. The key insight is to leverage techniques from secure multi-party computation to design protocols for decentralized computation in agent-based models. This ensures the confidentiality of the simulated agents without compromising on simulation accuracy. We showcase our protocols on a case study with an epidemiological simulation comprising over 150,000 agents. We believe this is a critical step towards deploying agent-based models to real-world applications.

\end{abstract}


\newcommand{\BibTeX}{\rm B\kern-.05em{\sc i\kern-.025em b}\kern-.08em\TeX}

\makeatletter
\gdef\@copyrightpermission{
	\begin{minipage}{0.3\columnwidth}
		\href{https://creativecommons.org/licenses/by/4.0/}{\includegraphics[width=0.90\textwidth]{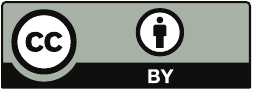}}
	\end{minipage}\hfill
	\begin{minipage}{0.7\columnwidth}
		\href{https://creativecommons.org/licenses/by/4.0/}{This work is licensed under a Creative Commons Attribution International 4.0 License.}
	\end{minipage}
	\vspace{5pt}
}
\makeatother

\begin{document}

%%% The following commands remove the headers in your paper. For final 
%%% papers, these will be inserted during the pagination process.

\pagestyle{fancy}
\fancyhead{}

%%% The next command prints the information defined in the preamble.

\maketitle 

%%%%%%%%%%%%%%%%%%%%%%%%%%%%%%%%%%%%%%%%%%%%%%%%%%%%%%%%%%%%%%%%%%%%%%%%

\input{content/introduction}
\input{content/abm}
\input{content/mpc}
\input{content/method}
\input{content/experiments}

\input{content/related_work}
\input{content/conclusions}

\begin{acks}
We are grateful to N. Bishop for his very insightful comments on the paper. This research was supported by a UKRI AI World Leading Researcher Fellowship awarded to M Wooldridge (grant EP/W002949/1). M. Wooldridge acknowledges funding from Trustworthy AI - Integrating Learning, Optimisation and Reasoning (TAILOR) (https://tailor-network.eu/), a project funded by European Union Horizon2020 research and innovation program under Grant Agreement 952215.
\end{acks}

%%%%%%%%%%%%%%%%%%%%%%%%%%%%%%%%%%%%%%%%%%%%%%%%%%%%%%%%%%%%%%%%%%%%%%%%

%%% The next two lines define, first, the bibliography style to be 
%%% applied, and, second, the bibliography file to be used.

\bibliographystyle{ACM-Reference-Format} 
\bibliography{paper}
%\bibliography{paper, paper2}

%\input{content/supplementary}

%%%%%%%%%%%%%%%%%%%%%%%%%%%%%%%%%%%%%%%%%%%%%%%%%%%%%%%%%%%%%%%%%%%%%%%%

\end{document}

%% file: content/introduction.tex
\section{Introduction}

% Outline comments

Agent-based modeling (ABM) is a bottom-up simulation technique wherein a system is modeled through the interaction of autonomous decision-making entities referred to as agents, which may represent individuals, companies, or other decision-making entities. Due to their granular approach, ABMs are a promising tool for real-world decision-making and policy design and constitute an active field of research across economics~\cite{axtellAgentBasedModelingEconomics, bonabeauAgentBasedModelingMethods2002, hollandArtificialAdaptiveAgents1991, carroHeterogeneousEffectsSpillovers2023}, biology~\cite{mordvintsevGrowingIsotropicNeural2022, gongComputationalMultiscaleAgentbased2017a}, and epidemiology~\cite{aylett-bullockJuneOpensourceIndividualbased2021a, romero2021public, kerrCovasimAgentbasedModel2021d, hinchOpenABMCovid19AgentbasedModel2021a}. The wider adoption of ABMs, however, is hindered by (1) the need for microdata to generate the underlying agent population and (2) the often substantial computational resources required to run, calibrate, and analyze an ABM. Recently, there has been significant progress towards developing new design patterns for ABMs, which exploit tensorization ~\cite{chopra2021deepabm, chopra2022decision} and differentiability ~\cite{chopraDifferentiableAgentbasedEpidemiology2023b, aryaAutomaticDifferentiationPrograms2022} of simulators. This has alleviated the computational burdens associated with ABM simulation~\cite{chopra2021deepabm}, calibration~\cite{chopraDifferentiableAgentbasedEpidemiology2023b, quera2023bayesian}, and analysis~\cite{queraDontSimulate} by granting access to techniques such as GPU computing and differentiable programming, allowing ABMs to scale to million-size populations \cite{quera-bofarullChallengesCalibratingDifferentiable2023, bhattacharya2023data}.

Unfortunately, improvements in the computational efficiency of ABMs is of little value if the quality of the underlying population microdata is poor. Currently, prevalent approaches involve the construction of synthetic populations designed to align with a predefined set of summary statistics derived from real-world observations. For instance, in epidemiological ABMs, the population is crafted to replicate summary statistics obtained from census data~\cite{mullerPopulationSynthesisMicrosimulation2010a, borysovHowGenerateMicroagents2019, aylett-bullockJuneOpensourceIndividualbased2021a, chapuisGenerationSyntheticPopulations2022a, pangalloUnequalEffectsHealtheconomy2022}. However, it is essential to recognize that the limited granularity of census data arises primarily from privacy considerations rather than the actual scarcity of available data. As ABMs continue to scale towards one-to-one representations of real-world systems, there remains a fundamental limitation in their modeling potential as long as privacy guarantees are not in place. Previous attempts to augment ABM data with additional information, such as mobility or health data, have resulted in data leaks that exposed agents' personal information \cite{IndonesiaProbesSuspected2021, kelleyIllinoisBoughtInvasive2021, coxTMobilePutMy2019}. These incidents underscore the need for a decentralized approach to ABM, where each agent's sensitive information is kept confidential throughout the modeling process.

Motivated by this, we introduce a new paradigm for agent-based simulation that ensures the confidentiality of each agent's sensitive information. Leveraging techniques drawn from secure multi-party computation \cite{lindell2020secure}, we develop privacy-preserving protocols for the simulation, calibration, and analysis of ABMs. These protocols offer robust security guarantees to agents while preserving the ability of ABMs to model complex systems effectively. Moreover, our methodology enables secure ABMs to take advantage of differentiable programming, allowing them to be integrated into machine learning pipelines, further boosting their modeling capabilities. We demonstrate the capabilities of this new methodology by running a case study in the city of Oxford, UK. We showcase how our approach can provide the same level of insight and analysis as traditional ABMs while guaranteeing the agents' privacy.
% \textcolor{red}{expts + impact of work..}.  

In summary, this work constitutes, to the best of our knowledge, the first framework for privacy-preserving ABMs. The framework supports their simulation, calibration, and analysis. We hope this development will pave the way for the secure and practical utilization of ABMs as valuable tools for policy-making in real-world settings.

%% file: content/abm.tex
\section{Agent-based models}
\label{sec:abms}

In this section, we formalize the processes of simulation, calibration, and analysis of ABMs. In doing so, we lay the foundations for the introduction of privacy-preserving protocols in \autoref{sec:decabm}.

\subsection{Simulation of ABMs}

Consider an ABM with $N$ agents $A = \{1, 2, \ldots, N\}$. We denote by $\mathbf z_i(t)$ the state of agent $i$ at time $t$, which encapsulates both fixed and time-evolving properties of the simulation agents. For example, $\mathbf z$ might represent the age and disease status of human agents in epidemiological models, or the account balance of firms in a financial auction model. As the simulation proceeds, an agent $i$ updates their state $\mathbf z_i(t)$ by interacting with their neighbors $\mathcal N_i(t)$ and the environment $\mathcal E(t)$. We assume that the interaction of agents with their neighbors can be conceived as message passing on a graph $\mathcal G=(V, E)$, where the vertices $V$ of the graph correspond to the agents,  edges $e_{ij}\in E$ connect neighboring agents, and interactions are represented as messages $M_{ij}(t) = M(\mathbf z_j(t), e_{ij}(t), \boldsymbol{\theta}, t)$, where $\boldsymbol{\theta}$ are the ABM structural parameters. This is the case for a diverse class of social and biological contagion models~\cite{doddsUniversalBehaviorGeneralized2004}. For example, $M_{ij}(t)$ may represent the transmission of infection from agent $j$ to agent $i$, which may depend on the susceptibility of agent $i$ ($\mathbf z_i$), the infectivity of $j$ ($\mathbf z_j$), the properties of the virus ($\boldsymbol{\theta}$), and the nature of the interaction ($e_{ij}$)~\cite{romero2021public, chopraDifferentiableAgentbasedEpidemiology2023b, hinchOpenABMCovid19AgentbasedModel2021a}; or transmission of information from agent $i$ to $j$ which depends on the opinion of agent $i$ ($\mathbf z_i$), and the assimilation and rejection thresholds of agent $j$ ($\mathbf z_j$)~\cite{coates2018unified, fotakis2016opinion, coates2018unified}. Thus, at step $t$, each agent updates its state as follows:
\begin{equation}
    \label{eq:agent_update}
    \mathbf z_i(t+1) =  f\left(\mathbf z_i(t), \bigoplus_{j \in \mathcal N_i(t)} M_{ij}(t), \; \boldsymbol{\theta}\right),
\end{equation}
where $\oplus$ denotes an aggregation function over all received messages. The specific form of $f$ can be tailored to capture the unique dynamics of the system under investigation. For instance, the diversity of contagion models can be encapsulated by different functional forms of $f$ \cite{doddsUniversalBehaviorGeneralized2004}.

During the simulation of an ABM, a central agent (the modeler) collects a time-series of aggregate statistics over agent states, $\mathbf x_t = h(\{\mathbf z_i(t) \mid i \in A\})$, which can be used to compare the output of the model to ground-truth data. For instance, in epidemiological ABMs, $h$ may correspond to counting the number of infected agents so that $\{\mathbf x_t\}_t$ is a time series of daily infections.

As we can see, both \autoref{eq:agent_update} and the collection of the summary statistics require agents to communicate their state to other agents. In the following sections, we introduce a methodology that enables these operations to take place while preserving the privacy of individual agents.

\subsection{Calibration of ABMs}
\label{sec:calibration}

Calibration refers to the process of tuning the set of structural parameters $\boldsymbol{\theta}$ so that ABM outputs $\mathbf x$ are compatible with given observational data $\mathbf y$. In epidemiological ABMs, for instance, this entails determining values for parameters like the reproduction number $R_0$ and mortality rates to align with the observed daily infection or mortality data. 

It is important to recognize that due to the stochasticity of the model and its partial observability, multiple sets of parameter values $\boldsymbol \theta$ may be compatible with the observed data $\mathbf{y}$. Consequently, it becomes essential to have an accurate  estimate of uncertainty around the calibrated parameters. Likewise, it is also important to be able to incorporate expert knowledge that may indicate a preference for certain regions of the parameter space over others into the calibration procedure. Both of these requirements can be met by adopting a Bayesian framework, wherein parameter inference corresponds to determining the posterior distribution over the parameters, $\pi(\boldsymbol{\theta} \mid \mathbf y)$ using Bayes' theorem,
\begin{equation}
    \pi (\boldsymbol{\theta} \mid \mathbf y) = \frac{p(\mathbf y\mid \boldsymbol{\theta}) \;\pi (\boldsymbol{\theta})}{p(\mathbf y)},
\end{equation}
where $\pi(\boldsymbol{\theta})$ is the prior distribution, $p(\mathbf y\mid \boldsymbol{\theta})$ is the likelihood function and $p(\mathbf y)$ is the marginal likelihood. For ABMs, the likelihood function is typically intractable; thus, we need to consider likelihood-free calibration algorithms.

While many Bayesian calibration methods exist for ABM (see, e.g., \cite{grazziniBayesianEstimationAgentbased2017b, plattComparisonEconomicAgentbased2020a, dyerBlackboxBayesianInference2022a}), we focus on methods that exploit the differentiability of the ABM. Differentiable ABMs \cite{andelfingerDifferentiableAgentBasedSimulation2021, chopraDifferentiableAgentbasedEpidemiology2023b} are ABMs implemented in frameworks that allow computing the gradient of the ABM output respect to the structural parameters, $\nabla_{\boldsymbol{\theta}} \mathbf x$, in an efficient way using techniques like automatic differentiation \cite{JMLR:v18:17-468}. Gradient-assisted calibration methods that take advantage of the differentiability of the ABM have been shown to be more efficient, scaling to larger parameter spaces than their gradient-free counterparts \cite{chopraDifferentiableAgentbasedEpidemiology2023b, quera-bofarullChallengesCalibratingDifferentiable2023}, without requiring the use of surrogate models. In most applications, the ABM output $\mathbf x$ is an aggregate over agent states through time, as is the case in epidemiology, where we are typically interested in infection curves. The gradient can then be computed as an aggregation
\begin{equation}
\label{eq:col_grad}
    \nabla_{\boldsymbol{\theta}} x_t = \bigoplus_{i\in A} \nabla_{\boldsymbol{\theta}} (h(\mathbf z_i(t))).
\end{equation}

A suitable likelihood-free approach to conduct gradient-assisted Bayesian calibration in ABMs is generalized variational inference \cite{knoblauch2022optimization} (GVI). In GVI, we employ a variational approach to target the generalized posterior,
\begin{equation}
\label{eq:genpost}
    \pi_{w}(\bth \mid \bd y) \propto e^{-w\cdot \ell(\bd{y}, \bth)} \pi(\bth),
\end{equation}
where $\ell(\bd{y}, \bth)$ is a loss function capturing the compatibility between the observed data $\bd{y}$ and the behaviour of the ABM at parameter vector $\bth$, and $w > 0$ is a hyperparameter. This posterior can then be approximated by finding a distribution $q$ in some variational family $\mathcal{Q}$ of distributions that minimises the Kullback-Liebler (KL) divergence to the generalised posterior given in \autoref{eq:genpost},
\begin{equation}
\label{eq:gvi_obj}
    q^* = \arg \min_{\boldsymbol \phi}{\mathcal L(\boldsymbol \phi)},
\end{equation}
where
\begin{equation}
    \mathcal L(\boldsymbol{\phi}) = \mathbb E_{q_{\boldsymbol{\phi}}} \left[ \ell(\mathbf x(\boldsymbol{\theta}), \mathbf y)\right] + w\; \mathrm{KL}(q_{\boldsymbol{\phi}}\mid\mid \pi(\boldsymbol{\theta})).
\end{equation}

This optimization problem can be tackled by estimating the gradient $\nabla_{\boldsymbol{\phi}}\mathcal L(\boldsymbol{\phi})$ and using an appropriate optimization technique such as stochastic gradient descent to minimize $\mathcal L(\boldsymbol{\phi})$. Given the differentiablity of the ABM, the gradient of the expected loss can be computed through a pathwise Monte Carlo gradient estimator  via the reparameterization trick (see \cite{mohamedMonteCarloGradient2020} for a comprehensive review),
\begin{equation}
    \label{eq:reparam}
    \nabla_{\boldsymbol{\phi}} \mathbb E_{q_{\boldsymbol{\phi}}} \left[ \ell(\mathbf x(\boldsymbol{\theta}), \mathbf y)\right] \approx \frac{1}{N}\sum_{i=1}^N \nabla_{\boldsymbol\phi} \ell (x(\boldsymbol \theta_{\boldsymbol \phi}(u^{(i)}), \mathbf y),
\end{equation}
where $u^{(i)} \sim p(u)$ is a sample from the base density $p(u)$ and $\bth_{\phi}(u^{(j)})$ is the transformed sample from the candidate posterior $q_{\phi}$. Finally, the gradient of the loss can be related to \autoref{eq:col_grad} via the chain-rule,
\begin{equation}
    \nabla_{\boldsymbol{\phi}}\ell(\mathbf x(\boldsymbol{\theta}, \mathbf y))= \nabla_{\boldsymbol{\phi}} \boldsymbol{\theta}_{\boldsymbol{\phi}} \cdot \nabla_{\boldsymbol{\theta}}\mathbf x.
\end{equation}

In practice, the variational family $\mathcal Q$ will be parameterized by a deep neural network (i.e, a normalizing flow) that is trained using the gradient in \autoref{eq:reparam}.

\subsection{Analysis of ABMs}
\label{sec:analysis}

One of the core strengths of ABMs is their granularity, which enables ABM to address research questions that are beyond reach for coarser methodologies. For instance, epidemiological ABMs can help analyze the observed disparities of COVID-19 infections among demographic groups \cite{khuntiEthnicityLinkedIncidence2020a, martinSociodemographicHeterogeneityPrevalence2020a, queraDontSimulate}, plan effective vaccination rollout plans \cite{zhouOptimizingSpatialAllocation2021}, or assess the role of latent transmission for a certain network structure \cite{marmorAssessingIndividualRisk2023}. All these studies require examination of the distribution of state values of the agents and their responsiveness to changes in the model's structural parameters. Specifically, we are interested in collecting a series of summary statistics over the agent's population, $\boldsymbol{\xi}_k = h_k(\{\mathbf z_i(t) \mid i \in A\})$, and their sensitivity, $\nabla_{\bth} \boldsymbol{\xi}_k$. For instance, in an epidemiological ABM, we might investigate the distribution of infections among different age groups and assess how these infections react to variations in the reproduction number $R_0$. Sensitivity analysis can be a challenging task due to the high computational cost of running large ABMs and the high dimensionality of the parameter space. As with the calibration process, differentiable ABMs may help mitigate this computational bottleneck \cite{queraDontSimulate}.

%% file: content/mpc.tex
\section{Characterizing Privacy}
First, we formalize a threat model to setup constraints for a privacy-preserving solution. Then, we provide background on secure multi-party computation and describe the GMW protocol which we use to design algorithms for privacy-preserving simulation, calibration and analysis of agent-based models.
% \arnau{This is incomplete?}

\subsection{Threat Model}
We assume an honest-but-curious (a.k.a.\ semi-honest) attacker~\cite{hazay2010note} which aims to learn private information about participating agents without altering the protocol. This private information is included in an agent's state $\mathbf z_{j}(t)$, interaction trace $\{\mathcal N_{i}(t) \mid \forall t\}$, and neighborhood messages $\{ M_{ij}(t) \mid i \in A, j \in \mathcal N(i)\}$. For instance, in epidemiological models, this can correspond to the health and demographic traits and mobility patterns of individual agents. Such an attacker can manifest itself as the coordinating server that wants to surveil agents using the mobility trace or a (sub-group) of adversarial agents, which may be incentivized to steal the personal health information of agent cohorts. In the context of agent-based modeling, this information can be leaked during message passing over per-step neighborhoods (\autoref{eq:agent_update}) and during the collection of summary statistics over the population. The goal of this work is to alleviate such challenges and design a privacy-preserving mechanism that can compute functions over agents' states without revealing private information. 

\subsection{Secure Multi-party Computation}
Secure multi-party computation enables a set of agents to interact and compute a joint function of their private inputs while revealing nothing but the output~\cite{lindell2020secure}. MPC protocols are coordinated with a server (MPC server) and are designed to protect against behavior of adversarial participants. These participants, either an agent or the server, aim to learn private information (of other entities) or cause the computation result to be incorrect. The idea was first introduced by Yao for the two-party case~\cite{yao1982protocols} and generalized to multiparty settings by Goldreich, Micali and Wigderson (GMW)~\cite{goldreich2019play}. Among other properties, GMW protocols guarantee (1) \textit{privacy}: so that no entity can learn anything more than its prescribed output and, (2) \textit{correctness}: so that agents receive the correct output. For instance, in an epidemiological ABM, this would ensure both that the personal disease status of agents is not leaked and that agents receive the correct transmission probability as in a centralized simulation.

% \giray{GMW doesn't guarantee correctness against malicious adversaries, an agent can misrepresent their disease status without the others figuring it out, that's partly why we need the semi-honest assumption}. \arnau{I think we should just say its 1) private and 2) accurate: assuming all agents are honest, the computation is accurate (as opposed to DP where noise may cause error)} 

\subsection{The GMW protocol}

The GMW protocol uses additive secret sharing to communicate (or aggregate) private inputs across the participant entities. The key insight is to divide a secret input into multiple shares in such a way that the secret can be reconstructed only when a sufficient number of shares are combined together. The scheme supports diverse aggregation queries such as secure addition or secure multiplication~\cite{beaver1992efficient} of the secrets held by the participating agents. Here we focus on the addition case and we assume that all participating agents are required to compute the secret, usually denoted by $t=N$, but the same methodology can be extended to multiplication and composite queries (see, e.g.,~\cite{lindell2020secure}).

Consider $N$ agents holding private values $s_i$. We want to compute the sum $\sum_i s_i$ without any agent $j$ acquiring knowledge about $s_{\{ k \neq j \}}$. To setup the protocol, the agents agree on an integer $n > \sum_i s_i$ defining the finite group $\mathbb Z_n$ on which all computations will be carried \footnote{The choice to perform finite group arithmetics is so that no information about the secret can be gained by holding $<N$ shares.}. Each agent $i$ then samples $N-1$ random numbers, $r_{ij} \sim \mathcal U\{0, n-1\}$, such that the input is divided into $N$ shares, $s_{ij}$ defined by
\begin{equation}
    s_i = \sum_{j=1}^N s_\mathrm{ij} \Mod n = \sum_{j=1}^{N-1} r_{ij} + \left(s_i - \sum_{j=1}^{N-1} r_{ij}\right) \Mod n.
\end{equation}
Each agent then sends each share of their secret to each corresponding agent; agent $i$ sends $s_{i1}$ share to agent 1, $s_{i2}$ share to agent 2, etc. Locally, each agent performs the sum
\begin{equation}
    \sigma_k = \sum_{i=1}^N s_{ik} \Mod n.
\end{equation}
Finally, all values $\sigma_k$ are shared so that the reconstructed sum,  $S = \sum_k\sigma_k \Mod n$, can be computed, corresponding to the sum of the agent inputs $s_i$ by construction. Typically, this reconstruction may be conducted by a central MPC server or a trusted agent. We summarize the protocol in \autoref{alg:ss}, and we provide an illustrating example below.

\subsubsection{Additive secret sharing example}
Consider $N= 3$ agents — Alice, Bob, and Carol — holding private values $s_A=2$, $s_B=3$, and $s_C=5$. They wish to compute the sum of these values without disclosing their individual inputs. They agree on an integer $n=11$, defining a finite group $\mathbb Z_n$. First, the agents generate 3 shares each by sampling 2 random numbers from $\mathbb Z_n$. For instance, Alice generates random numbers 7 and 5, so that 
\begin{equation}
    s_A = s_{AA} + s_{AB} + s_{AC} = 7 + 5 + 1 \Mod{11} = 2,
\end{equation}
and similarly for Bob and Carol with $s_B = 2 + 0 + 1 \Mod{11}$, and $s_C = 3 + 1 + 1 \Mod{11}$. Second, the agents communicate with each other to keep one of the shares and send the other two to the other two agents and perform the sum of the received shares. For example, Alice receives $s_{BA}$ from Bob and $s_{CA}$ from Carol and computes 
\begin{equation}
    \sigma_A = s_{AA} + s_{BA} + s_{CA} = 7 + 2 + 3 \Mod {11} = 1 \Mod {11},
\end{equation}
and similarly for Bob and Carol with $\sigma_B = 5 + 0 + 1 \Mod{11} = 6 \Mod {11}$ and $\sigma_C = 1 + 1 + 1 \Mod {11} = 3 \Mod {11}$. 
Finally, the secret can be reconstructed by doing $S = \sigma_A + \sigma_B + \sigma_C = 10 \Mod{11}$ as expected. 

In the following section, we apply the GMW protocol to generalize the above insight to share information containing agent's private information to other agents or a central MPC server, providing protocols for the computation of agent updates (\autoref{eq:agent_update}), and gradients (\autoref{eq:col_grad}) in a secure way, enabling privacy-preserving simulation, calibration, and analysis of ABMs.

\begin{algorithm}[h]
\caption{\textsc{SecureSum}}
\label{alg:ss}
\KwData{Agents $\{1, \ldots, N\}$ with secret inputs $s_1, \ldots, s_n$, integer $n > \sum_i s_i $.}
\KwResult{The sum of all shares $S = s_1 + \cdots + s_n$.}

\textbf{Splitting secret into shares and distributing:}\\
Each party $i$ generates $N$ shares $s_{i1}, \ldots, s_{iN} \in \Z_n$ which sum up to $s_i$.

Each party $i$ distributes all their shares $s_{i1}, \ldots, s_{iN} \in \Z_n$ to $1, \ldots, N$, including themselves.

\textbf{Secure Computation (Addition):}\\
To add the inputs securely, parties simply add their respective shares $\sigma_i = s_{1i} + \cdots s_{Ni} \mod n$.

\textbf{Reconstruction:}\\
To reveal the final result of the computation, parties collaborate by summing their shares:\\
$S = (\sigma_1 + \sigma_2 + \cdots + \sigma_n) \mod n$.\\
\end{algorithm}

%\begin{figure}[h]
%\begin{tikzpicture}
%  \pgfmathsetmacro{\w}{0.3}
%  \node (person1) at (0, -1) [label={[align=center, draw]below:$\sigma_A = s_{AA} + s_{BA} + s_{CA}$}]{\includegraphics[width=\w cm]{figures/user.png}};
%  \node (person2) at (-2.5, 1.5) [label={[align=center, draw]above:$\sigma_B = s_{AB} + s_{BB} + s_{CB}$}]{\includegraphics[width=\w cm]{figures/user.png}};
%  \node (person3) at (2.5, 1.5) [label={[align=center, draw]above:$\sigma_C = s_{AC} + s_{BC} + s_{CC}$}]{\includegraphics[width=\w cm]{figures/user.png}};
%  
%  \draw[->] (person1) to[bend right=10] node[sloped, align=center, above] {$\scriptstyle s_{AB}$} (person2);
%  \draw[->] (person2) to[bend right=10] node[sloped, align=center, below] {$\scriptstyle s_{BA}$} (person1);
%  \draw[->] (person1) to[bend right=10] node[sloped, align=center, below] {$\scriptstyle s_{AC}$} (person3);
%  \draw[->] (person3) to[bend right=10] node[sloped, align=center, above] {$\scriptstyle s_{CA}$} (person1);
%  \draw[->] (person2) to[bend right=10] node[sloped, align=center, below] {$\scriptstyle s_{BC}$} (person3);
%  \draw[->] (person3) to[bend right=10] node[sloped, align=center, above] {$\scriptstyle s_{CB}$} (person2);
%
%\end{tikzpicture}
%  \caption{Diagram describing the communication of shares between parties of the described example for the GMW protocol (\autoref{alg:ss}) with 3 agents.}
%  \label{fig:decabm_diagram}
%\end{figure}

%% file: content/method.tex
% \section{The Agent-based Model}
% Describe epidemiological transmission and progression model in a centralized setting.

\section{Private Agent-based Models}
\label{sec:decabm}

With the MPC background introduced in the preceding section, we formulate protocols to conduct the simulation, calibration, and analysis of ABMs in a privacy-preserving way.

\subsection{Secure Simulation}
\label{sec:secure_simulation}

In order to update the state of an agent during a simulation, \autoref{eq:agent_update} requires an aggregation over the agent's neighbors, revealing their private information. As described in the previous section, MPC is well-equipped to perform this kind of calculation without revealing the individual parties' data. Without loss of generality, we present our protocols for the case where the aggregation function $\oplus$ is a summation $\Sigma$, so that we can make use of the \textsc{SecureSum} protocol introduced in \autoref{alg:ss}. Furthermore, as long as the agent's update function $f$ is differentiable respect to the structural parameters $\boldsymbol{\theta}$, which is indeed the case for many ABMs \cite{chopraDifferentiableAgentbasedEpidemiology2023b}, each agent can store $\nabla_{\boldsymbol{\theta}} f$ for use during the calibration step. With all this in mind, we present in \autoref{alg:state}, a privacy-preserving protocol for updating agent states.

\begin{algorithm}[h]
\caption{\textsc{SecureAgentUpdate}}
\label{alg:state}
\KwData{Agent $i$ with state $\mathbf z_i(t)$, Neighboring agent's messages $\{M_{ij}(t) \mid j \in \mathcal N(i)\}$, Integer $n$, State update rule $f$, ABM parameters $\boldsymbol{\theta}$}
\KwResult{New state $\mathbf z_i(t+1)$}

Agent $i$ calls the \textsc{SecureSum} protocol with neighbors $\{j \mid j \in \mathcal N(i)\}$ and integer $n$ to obtain the sum $M_i(t) = \sum_{j \in\mathcal N(i)} M_{ij}(t)$.

Agent $i$ updates its state $\mathbf z_i(t+1) = f\left(\mathbf z_i(t), M_i(t), \boldsymbol{\theta} \right)$ and stores the gradient $\nabla_{\boldsymbol{\theta}} f$.

\end{algorithm}

It is worth noting that, in contrast to general applications of the GMW protocol, only the agent who starts the protocol receives the result of the computation since there is no need for the neighboring agents to have access to that information.

Next, we introduce the \textsc{SecureSimulation} protocol in \autoref{alg:simulation}, where, in addition to performing agent updates, we collect a time series of aggregate statistics over the agent's population and its gradient with respect to the ABM structural parameters $\boldsymbol{\theta}$.

\begin{algorithm}[h]
\caption{\textsc{SecureSimulation}}
\label{alg:simulation}

\KwData{MPC server $C$, Agents $\{1, \ldots, N\}$ with states $\{\mathbf z_1, \ldots, \mathbf z_N\}$, ABM parameters $\boldsymbol{\theta}$, State update rule $f$, Number of time-steps $T$}
\KwResult{Aggregate statistics $\mathbf x = x_1, \ldots, x_T$ and gradients $\nabla_{\boldsymbol{\theta}} \mathbf x$.}

$C$ generates a large enough prime number $P$  and the requested statistics collecting function $h$; and sends them to all agents along ABM parameters $\boldsymbol{\theta}$.

\For{t = 1, \ldots, T}{
    \For{i = 1, \ldots, N}{
        Agent $i$ calls the \textsc{SecureAgentUpdate} protocol (\autoref{alg:state}) to compute $\mathbf z_i(t+1)$.
        
        Agent $i$ gathers its information of interest $h(\mathbf z_i(t+1))$ and gradient $\nabla_{\boldsymbol{\theta}} h(\mathbf z_i(t+1))$.
    }
    
    $C$ calls the \textsc{SecureSum} protocol with all agents to collect the aggregate statistics $x_t$ and their gradients $\nabla_{\boldsymbol{\theta}} x_t$.
}
    $C$ returns the accumulated $\mathbf x$ and $\nabla_{\boldsymbol{\theta}}\mathbf x$.
\end{algorithm}

\begin{figure}
\begin{tikzpicture}[scale=0.6]
  \pgfmathsetmacro{\w}{0.3}
  \node (theta) at (-2, 2) {\Large $\boldsymbol \theta$};
  \node (person1) at (1, 0) {\includegraphics[width=\w cm]{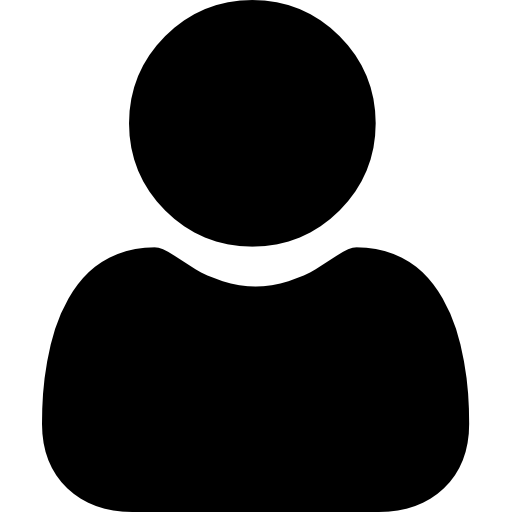}};
  \node (person2) at (0.6, 1.4) {\includegraphics[width=\w cm]{figures/user.png}};
  \node (person3) at (1.0, 4) {\includegraphics[width=\w cm]{figures/user.png}};
  \node (person4) at (1.7, 2.5) {\includegraphics[width=\w cm]{figures/user.png}};

  \node (grad1) at (2.2, 0) [align=center] {\Small $h(\mathbf z_1)$ \\ \Small $\nabla_{\boldsymbol \theta} h(\mathbf z_1)$};
  \node (grad2) at (1.7, 1.4) [align=center] {\Small $h(\mathbf z_2)$ \\ \Small $\nabla_{\boldsymbol \theta} h(\mathbf z_2)$};
  \node (grad3) at (2.0, 4) [align=center] {\Small $h(\mathbf z_3)$ \\ \Small $\nabla_{\boldsymbol \theta} h(\mathbf z_3)$};
  \node (grad4) at (2.8, 2.5) [align=center] {\Small $h(\mathbf z_4)$ \\ \Small $\nabla_{\boldsymbol \theta} h(\mathbf z_4)$};
  %\node (grad5) at (3.2, 1.3-0.5) [draw,thick] {$\nabla_\theta f_5$};
  
  \node (ss) at (6.5, 2) {\textsc{SecureSum}};

  \node (grad) at (6.5, 0.5) [draw, thick, align=center] {\Small $x_t = \sum_i h(\mathbf z_i)$ \\ $\nabla_{\boldsymbol{\theta}} x_t = \sum_i \nabla_{\boldsymbol{\theta}} h(\mathbf z_i)$};
  
  \draw[dashed] (person1) -- (person2);
  \draw[dashed] (person2) -- (person3);
  \draw[dashed] (person2) -- (person4);
  \draw[dashed] (person3) -- (person4);
  %\draw[dashed] (person4) -- (person5);

  \draw[->] (theta) -- (person1);
  \draw[->] (theta) -- (person2);
  \draw[->] (theta) -- (person3);
  \draw[->] (theta) -- (person4);
  %\draw[->] (theta) -- (person5);

  \draw[->] (grad1) -- (ss);
  \draw[->] (grad2) -- (ss);
  \draw[->] (grad3) -- (ss);
  \draw[->] (grad4) -- (ss);
  %\draw[->] (grad5) -- (ss);

  \draw[->] (ss) -- (grad.north);
\end{tikzpicture}
  \label{fig:grad_tikz}
  \caption{Diagram illustrating the \textsc{SecureSimulation} protocol for ABM parameters $\boldsymbol{\theta}$.}
\end{figure}

\subsection{Secure Calibration}
\label{sec:secure_calibration}

During the calibration of an ABM, the modeler (central MPC server) requires the ability to evaluate the ABM at different values of $\boldsymbol{\theta}$, and, in the case of gradient-assisted calibration, the gradient of the outputs with respect to $\boldsymbol{\theta}$. The retrieval of these quantities is enabled by the \textsc{SecureSimulation} protocol and so classical calibration algorithms \cite{grazziniBayesianEstimationAgentbased2017b, plattComparisonEconomicAgentbased2020a} can be seamlessly applied to conduct a privacy-preserving calibration. Here, as outlined in \autoref{sec:calibration}, we focus on conducting GVI with a deep neural network trained to approximate the generalized posterior. This neural network seats on the central MPC server and it is trained using the collected gradients. To this end, what remains to be detailed is the transition from the gradient of aggregate statistics, $\nabla_{\boldsymbol{\theta}} \mathbf{x}$, to the loss gradient required for GVI (\autoref{eq:reparam}). This process is detailed in \autoref{alg:calibration}. 

\begin{algorithm}[h]
\caption{\textsc{SecureCalibration}}
\label{alg:calibration}

\KwData{MPC server $C$, Agents $\{1, \ldots, N\}$ with states $\{\mathbf z_1, \ldots, \mathbf z_N\}$, State update rule $f$, Prior $\pi(\boldsymbol\theta)$, Observed time-series $\mathbf y$, Loss function $\ell(\mathbf x, \mathbf y)$, Number of epochs $N_e$, Number Monte-Carlo samples $N$, Number of time-steps $T$, Learning rate $\eta$}
\KwResult{Trained candidate posterior $q^*$}

$C$ initializes candidate posterior $q_{\boldsymbol{\boldsymbol \phi}}$.

\For {1, \ldots, $N_e$}{
    \For {1, \ldots, N}{
    
    $C$ samples ABM parameters $\boldsymbol{\theta}\sim q_{\boldsymbol \phi}(\boldsymbol{\theta})$.

    $C$ executes \textsc{SecureSimulation} protocol (\autoref{alg:simulation}) to obtain $\mathbf x$ and $\nabla_{\boldsymbol\theta} \mathbf x$.

    $C$ uses the chain rule to compute 
    \begin{equation*}
        \nabla_{\boldsymbol\theta}\ell(\mathbf x(\boldsymbol \theta, \mathbf y)) = \nabla_{\mathbf x}\ell(\mathbf x, \mathbf y) \cdot \nabla_{\boldsymbol{\theta}} \mathbf x
    \end{equation*}
    \begin{equation*}
    \nabla_{\boldsymbol\phi} \ell (\mathbf x(\boldsymbol\theta), \mathbf y) = \nabla_{\boldsymbol\theta}\ell(\mathbf x(\boldsymbol \theta), \mathbf y) \cdot \nabla_{\boldsymbol{\phi}}\boldsymbol{\boldsymbol\theta}
    \end{equation*}
    using the reparameterization trick (\autoref{eq:reparam}).
    }

    $C$ accumulates gradient to estimate
    \begin{equation*}
    \nabla_{\boldsymbol{\phi}} \mathbb E_{q_{\boldsymbol{\phi}}} \left[ \ell(\mathbf x(\boldsymbol{\theta}), \mathbf y)\right] \approx \frac{1}{N}\sum_{i=1}^N \nabla_{\boldsymbol\phi} \ell (x(\theta), y).
    \end{equation*}

    $C$ computes divergence $\mathrm{KL}(q_{\boldsymbol{\phi}}\mid\mid \pi(\boldsymbol{\theta}))$ and derivative $\nabla_{\boldsymbol{\phi}} \mathrm{KL}(q_{\boldsymbol{\phi}}\mid\mid \pi(\boldsymbol{\theta}))$.

    $C$ updates $\boldsymbol \phi \rightarrow \boldsymbol \phi - \eta \nabla{\boldsymbol{\phi}}\mathcal L(\boldsymbol{\phi})$, where 
    \begin{equation*}
        \mathcal L(\boldsymbol{\phi}) = \mathbb E_{q_{\boldsymbol{\phi}}} \left[ \ell(\mathbf x(\boldsymbol{\theta}), \mathbf y)\right] + \mathrm{KL}(q_{\boldsymbol{\phi}}\mid\mid \pi(\boldsymbol{\theta})).
    \end{equation*}
}
    $C$ returns trained candidate posterior $q^*$.
\end{algorithm}

\subsection{Secure Analysis}
\label{sec:secure_analysis}

As outlined in \autoref{sec:analysis}, we require a secure protocol to retrieve summary statistics over the agent's population, $\boldsymbol{\xi}_k = h_k(\{\mathbf z_i(t) \mid i \in A\})$, and their sensitivity, $\nabla_{\bth} \boldsymbol{\xi}_k$. We note that we have already addressed this issue in the \textsc{SecureSimulation} protocol since retrieving the time-series $\mathbf x$ is a particular case of this more general problem. In fact, we formulate a more general approach wherein a summary statistic $\boldsymbol{\xi}$ over the population states can be obtained by aggregating over the entire population the outputs of an indicator function $\mathbbm 1_{\Omega}(\mathbf z)$ acting on the agent's state,
\begin{equation}
    \xi = \bigoplus_{i \in A} \mathbbm 1_{\Omega}(\mathbf z_i),
\end{equation}
where $\Omega$ denotes the set of characteristics we want to aggregate on, and $\oplus$ denotes a kind of aggregation, usually a sum. For instance, $\Omega$ may correspond to the property of being infected, in which case taking $\oplus$ to be a sum would return the number of infected agents.
The algorithm is formally described in \autoref{alg:analysis}, which holds the same security guarantees as the one exposed in \autoref{sec:secure_simulation}.

\begin{algorithm}[h]
\caption{\textsc{SecureSummaryStatistic}}
\label{alg:analysis}

\KwData{MPC server $C$, Agents $\{1, \ldots, N\}$ with states $\{\mathbf z_1, \ldots, \mathbf z_N\}$, Indicator function $\mathbbm 1_{\Omega}(\mathbf z)$.}
\KwResult{Aggregate quantity $\bigoplus_{i\in A} \mathbbm 1_{\Omega}(\mathbf z_i)$.}

$C$ generates prime number $P$ and sends them to all agents alongside indicator function $\mathbbm 1_{\Omega}(\mathbf z)$.

Each agent $i$ computes result of $\mathbbm 1_{\Omega}(\mathbf z_i)$.

$C$ executes \textsc{SecureSum} protocol to securely retrieve $\bigoplus_i^N \mathbbm 1_{\Omega}(\mathbf z_i)$.

\end{algorithm}

Likewise, sensitivity analysis can be performed securely using \autoref{alg:sensitivity}, where each agent computes the sensitivity of their state change to the model parameters, and the central agent retrieves the aggregation by employing the \textsc{SecureSum} protocol.

\begin{algorithm}[h]
\caption{\textsc{SecureSensitivityAnalysis}}
\label{alg:sensitivity}

\KwData{MPC server $C$, Agents $\{1, \ldots, N\}$ with states $\{\mathbf z_1, \ldots, \mathbf z_N\}$, Indicator function $\mathbbm 1_{\Omega}(\mathbf z)$, State update rule $f$}
\KwResult{Sensitivity $\nabla_{\boldsymbol{\theta}} \mathbf \xi$.}

$C$ generates prime number $P$ and sends them to all agents alongside indicator function $\mathbbm 1_{\Omega}(\mathbf z)$.

Each agent $i$ executes \textsc{SecureAgentUpdate} with parameters $\boldsymbol{\theta}$ to obtain $\nabla_{\boldsymbol{\theta}} f(\mathbf z_i, M_i, \boldsymbol{\theta})$ so that
$$
\nabla_{\boldsymbol \theta}\xi_i = \nabla_{\boldsymbol{\theta}} f(\mathbf z_i, M_i, \boldsymbol{\theta}) \cdot \mathbbm 1_{\Omega}(\mathbf z)
$$

$C$ executes \textsc{SecureSum} protocol to securely retrieve $\nabla_{\boldsymbol{\theta}} \mathbf \xi = \bigoplus_i^N \nabla_{\boldsymbol \theta}\xi_i$.

\end{algorithm}

%% file: content/experiments.tex
\section{Case Study: Privacy-preserving Epidemiology}

In this section, we illustrate how our methodology may be deployed in practice by describing a decentralized privacy-preserving agent-based SIR model, which supports simulation, calibration, and analysis.

The model follows a standard parameterization where agents' interactions are specified through a contact graph $\mathcal G$, which in this case is only locally defined by each agent having access to their neighbors. Each agent has 3 possible states: 0 (Susceptible), 1 (Infected), and 2 (Recovered). We initialize the simulation by infecting a fraction $I_0$ of agents sampled uniformly from the population, while the remaining agents are considered susceptible. Following the notation introduced in \autoref{sec:abms}, at each time-step, agent $i$ updates its state following \autoref{eq:agent_update} with
\begin{equation}
    M_{ij}(t) = I_j(t) 
\end{equation}
where $I_j(t)$ is the infected status of the neighbor (0 or 1), so that
\begin{equation}
\begin{split}
    z_i(t+1) = \; &\mathbbm 1_{\{z_i = 0\}} \cdot \mathrm{Bernoulli} \left(p_\mathrm{inf}^{(i)}(t)\right)  \; +\\
    &\mathbbm 1_{\{z_i = 1\}} \cdot \left(1 + \mathrm{Bernoulli}\left(p_\mathrm{rec}^{(i)}\right)\right) \;  + \\
    & \mathbbm 1_{\{z_i =2\}} \cdot  2
\end{split}
\end{equation}
with
\begin{equation}
    \label{eq:infection}
    p_\mathrm{inf}^{(i)}(t) = 1 - \exp\left(-\frac{\beta\, S_i \Delta t}{n_i} \sum_{j \in \mathcal N(i)} I_j(t)\right),
\end{equation}
where $\mathcal N(i)$ is the set of neighbors of agent $i$, $S_i$ is the susceptibility of agent $i$, $n_i = \# \mathcal N(i)$ is the total number of neighbors, $\Delta t$ is the duration of the time-step, and $\beta$ is a structural parameter of the ABM called the effective contact rate. Infected agents can recover at each time step with recovery rate $\gamma$, so that
\begin{equation}
    p_\mathrm{rec}^{(i)} = 1 - \exp\left(-\gamma \Delta t\right).
\end{equation}
For the case of a complete graph, the model reduces to the standard ODE-based SIR model with $R_0 = \beta/\gamma$ as the basic reproduction number. The model is run for $n_t$ time steps.

To ground the example on real data, we consider the contact graph of the city of Oxford, extracted from the June ABM model \cite{aylett-bullockJuneOpensourceIndividualbased2021a} to determine the neighborhood of each agent, $\mathcal N(i)$. This contact graph includes agents' interactions in households, companies, and schools, and it is based on English census data. The choice of parameter values for the experiment is given in \autoref{tab:sir_parameters}.

\begin{table}[h]
    \centering
    \begin{tabular}{cl}
         Parameter & Value\\
         \hline
         $\beta$ & $0.5 \; \mathrm{day}^{-1}$\\
         $\gamma$ & $0.1 \; \mathrm{day}^{-1}$\\
         $I_0$ & 0.01 \\
         $\Delta t$& 1 day \\
         $n_t$& 60 \\
         $\mathcal G$ & Oxford\\
    \end{tabular}
    \caption{Parameter values for the agent-based SIR model.}
    \label{tab:sir_parameters}
\end{table}

\subsection{Private policy assessment with ABMs}

\begin{figure}[h]
    \includegraphics[width=0.7\columnwidth]{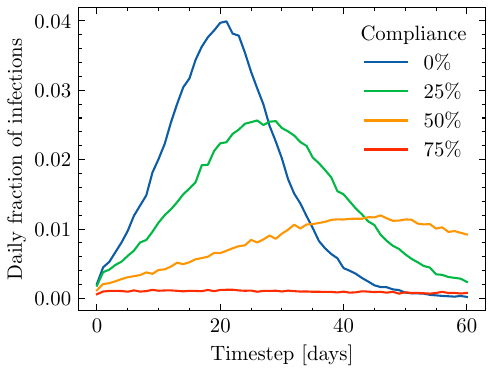}
    \caption{Infection curves for different levels of compliance: 0\% (blue), 25\% (green), 50\% (orange), 75\% (red). The number of infections has been normalized to the number of agents $N$. These plots are generated without releasing the infection status or compliance decision of any individual agent.}
    \label{fig:compliance}
\end{figure}

We first consider the application of the \textsc{SecureSimulation} protocol (\autoref{alg:simulation}). Let us pose a situation where a policymaker wants to study the efficacy of mask-wearing at different compliance levels using agent-based simulation. We introduce a slight modification to \autoref{eq:infection} to incorporate a reduction in the infection probability due to mask-wearing with certain compliance $\alpha$,
\begin{equation}
    \label{eq:compliance}
    p_\mathrm{inf}^{(i)}(t) = 1 - \exp\left(-\frac{\beta\, S_i\Delta_t}{n_i} \sum_{j \in \mathcal N(i)} I_j(t) (1-c_j)\right),
\end{equation}
where $c_j \sim \mathrm{Bernoulli}(\alpha)$, so that $\alpha_i=1$ corresponds to full compliance where there is no transmission. Note that we are assuming complete protection against infection when wearing a mask. We proceed to execute 3 simulations for 3 different values of $\alpha$. At each simulation, $\alpha$ is sent to the agents, where they locally compute their own compliance to the measure. The \text{SecureSimulation} protocol is then used to run the simulation and retrieve the aggregate statistic of interest, $\mathbf x$, which in this case is the number of infections over time. The results are shown in \autoref{fig:compliance}, where we observe that little transmission occurs when compliance is above $\gtrapprox 75\%$.

Thus, we observe that within our methodology, the policymaker could still have access to the same level of insight as a traditional ABM, all while protecting the individual agent's privacy.

\subsection{Private calibration of ABMs}
\label{sec:experiment_calibration}

Next, we pose a situation where we want to calibrate our ABM with structural parameters $\boldsymbol{\theta} = (\beta, \gamma)$ to observed ground-truth data. For simplicity, we present the calibration of the $\beta$ parameter given an observed curve of infections ($\mathbf y$), obtained by running the ABM model with the baseline parameters in \autoref{tab:sir_parameters}.

\begin{figure}
     \centering
     \begin{subfigure}[b]{0.4\columnwidth}
        \includegraphics[width=\columnwidth]{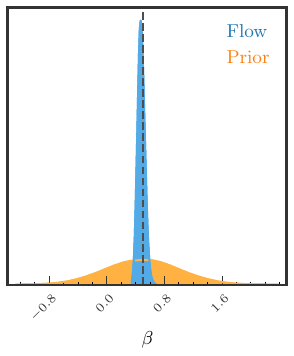}
     \end{subfigure}
     \begin{subfigure}[b]{0.55\columnwidth}
        \includegraphics[width=\columnwidth]{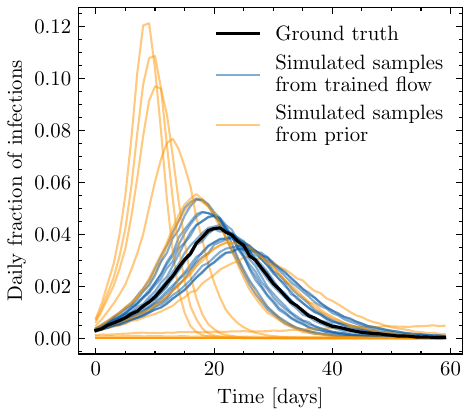}
     \end{subfigure}
        \caption{Left: Probability density plot for the trained normalizing flow (blue) against the prior distribution (orange). Ground-truth value is marked as a dashed black line. Right: Results from simulating $\beta$ samples from the trained flow (blue) and prior (orange) compared to the ground-truth data (black). The number of infections has been normalized to the number of agents $N$.}
        \label{fig:trained_flow}
\end{figure}

The first step is to compute the gradient $\nabla_{\boldsymbol{\theta}} \mathbf{x}$, where $\mathbf x$ is the number of daily infections and $\theta = \beta$. We note that this gradient can be approximated by the gradient of the average number of new infections with respect to $\beta$,
\begin{equation}
    \frac{\partial x_t}{\partial\beta} \approx \frac{\partial \; \mathbb{E}[\Delta I(t)]}{\partial \beta} = \sum_{i=1}^N \chi_i(t) \exp(-\chi_i(t) / \beta),
\end{equation}
where
\begin{equation}
    \chi_i(t) = \exp\left(-\frac{\beta\, S_i\Delta_t}{n_i} \sum_{j \in \mathcal N(i)} I_j(t) \right).
\end{equation}
A central agent can safely retrieve the gradient by performing the \textsc{SecureSum} protocol across all agents as described in \autoref{alg:simulation} and \autoref{alg:calibration}. We thus conduct GVI by considering $\mathcal Q$ to be a masked-autoregressive normalizing flow \cite{papamakariosMaskedAutoregressiveFlow2017} and assume the prior is a normal distribution with $\mu=0.7$ and $\sigma=0.5$. 
\autoref{fig:trained_flow} (left) shows the trained normalizing flow, which correctly assigns high probability mass to the ground-truth value. To further evaluate the goodness of the fit, we plot simulated runs from ABM parameters sampled from the trained flow in \autoref{fig:trained_flow} (right), where we compare it to runs simulated from prior samples. 

This experiment highlights how privacy-preserving ABMs can be integrated into differential and probabilistic programming pipelines. This opens the door to integrating ABM insight into more complex ML pipelines leveraging heterogeneous data streams to boost the model's insight capabilities.

\subsection{Private demographic study with ABMs}

In this section, we apply the \textsc{SecureSummaryStatistic} protocol to analyze our calibrated ABM. In particular, we study the distribution of infections by age, ethnicity, and geographical location (ZIP code). This analysis may be relevant to, for instance, understanding the causes of asymmetric distribution of infections among different demographic groups ~\cite{khuntiEthnicityLinkedIncidence2020a, martinSociodemographicHeterogeneityPrevalence2020a, queraDontSimulate}.

For this exercise, we import a synthetic population of Oxford from the June model, as we did for the contact graph, so that we have access to a population with realistic age, sex, and ethnicity distributions and geographical location. Note that the ethnicity categorization follows the English census \cite{gov.ukListEthnicGroups}. 

\begin{figure}[h]
    \includegraphics[width=\columnwidth]{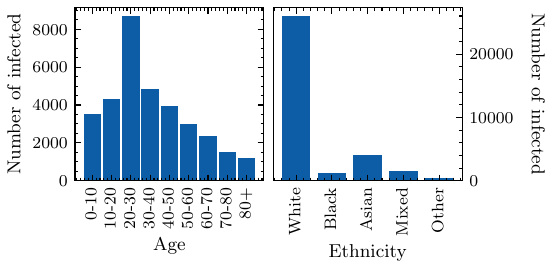}
    \caption{Age (left) and ethnicity (right) histogram of the infected population. These statistics are computed without leaking the infection or demographic properties of any agent.}
    \label{fig:inf_histogram}
\end{figure}

We can construct histograms that highlight the distribution of infection among different ethnic groups by considering the relevant indicator functions so that we can apply the \textsc{SecureSummaryStatistic} protocol. For instance, for the case of age distribution, we have
\begin{equation}
    \mathbbm 1_\Omega (\mathbf z) = \mathbbm 1_{\{(z_\mathrm{age} \in a_k) \land (z_\mathrm{inf} = 1)\}}(\mathbf z), 
\end{equation}
where $z_\mathrm{age}$ and $z_\mathrm{inf}$ are the age and infected status of the agent, and $a_k$ are each of the histogram age bins. The number of counts in the particular age bin $a_k$ can then be obtained upon executing \textsc{SecureSum} over the entire population following \autoref{alg:analysis}. \autoref{fig:inf_histogram} shows the age and ethnicity histogram of infections for the calibrated simulation obtained in \autoref{sec:experiment_calibration}, where the ethnicity histogram has been computed analogously to age one by considering the indicator function 
\begin{equation}
    \mathbbm 1_\Omega(\mathbf z) = \mathbbm 1_{\{(z_\mathrm{ethnicity} = e_i) \land (z_\mathrm{inf} = 1)\}}(\mathbf z), 
\end{equation}
for each ethnicity category $e_i$. The results are in agreement with the particular demographics of the city of Oxford, dominated by students in the 20-30 age bin with a significant presence of non-White groups.

\begin{figure}[h]
    \includegraphics[width=0.8\columnwidth]{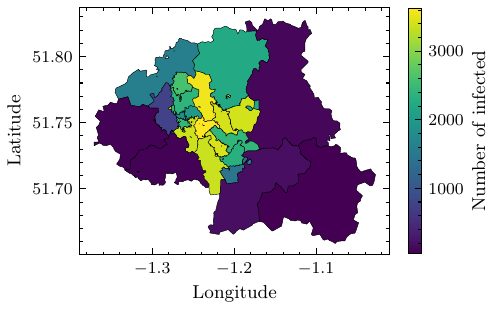}
    \caption{Geographical distribution of infections by ZIP code  sector within the city of Oxford. These statistics are computed without leaking the infection status or geo-location of any individual agent.}
    \label{fig:geo_infected}
\end{figure}

Finally, a similar analysis can be done at the geographical level, where infections can be collected by ZIP code sector by executing the \textsc{SecureSummaryStatistic} protocol with $\mathbbm 1_{\Omega(\mathbf z)} = \mathbbm 1 _{\{z_\mathrm{ZIP code} = r_i\}}$ for each ZIP code sector $r_i$. In \autoref{fig:geo_infected}, we show the obtained distribution of infections across the city of Oxford.

Summarising, we have illustrated how a detailed analysis of agent properties can be performed in a privacy-preserving way by following \autoref{sec:secure_analysis}.

%% file: content/related_work.tex
\section{Related Work}
Before concluding, we survey recent work on data-driven ABMs, with a particular focus on recent advances that aim to deploy ABMs to real-world populations with millions of agents. We also highlight the privacy challenges that arise from extending such ABMs.

% \textcolor{red}{mention why not for GNNs}

\subsection{Data-driven Agent-based Modeling}
The push to create highly realistic ABMs has led to a growing demand for more granular heterogeneous data to construct the synthetic populations that underlie these models. This demand is particularly evident in fields such as epidemiology (e.g., \cite{kerrCovasimAgentbasedModel2021d, aylett-bullockJuneOpensourceIndividualbased2021a, chopra2021deepabm}) and economics (e.g., \cite{carroHeterogeneousEffectsSpillovers2023, polednaEconomicForecastingAgentbased2023}), where the population is constructed from census data, using techniques such as iterative proportional fitting \cite{ruschendorfConvergenceIterativeProportional1995, choupaniPopulationSynthesisUsing2016} and deep generative modelling  \cite{borysovHowGenerateMicroagents2019}. Furthermore, the integration of dynamic data streams, including sources like SafeGraph mobility data, CDC genomic data, and Facebook survey data, into the ABM calibration process has become essential for generating real-time insights. In addressing this need, the combination of differentiable ABMs and deep neural networks (DNN) stands out as an effective approach to seamlessly incorporate heterogeneous data sources during the calibration process \cite{chopraDifferentiableAgentbasedEpidemiology2023b}. Extending beyond the current resolution of ABM populations requires the development of new methodologies that safeguard the privacy of stakeholders, which is the objective of this work.

\subsection{Data Privacy for Modeling}

% \textcolor{red}{a) privacy methods and risk for ABM data. b) privacy techniques in other disciplines: also mention GNN and MPC protocols exist but more challenging. Viewing ABMs as message passing lens allows to utilize that. c) MPC for modeling for contact tracing and CSAM. No privacy-preserving calibration algorithms. We give unified framework for all.}

%The aforementioned advances in ABM computation have allowed for the simulation and calibration of populations comprised of millions of agents. However, the modeling of one-to-one populations carries an associated need for more granular heterogeneous data. Contributions aimed at addressing the issue include novel approaches for generating synthetic populations from census data, such as iterative proportional fitting \cite{ruschendorfConvergenceIterativeProportional1995, choupaniPopulationSynthesisUsing2016} and deep generative modelling  \cite{borysovHowGenerateMicroagents2019}. Often, multiple heterogeneous data sources need to be combined. For instances,  SafeGraph mobility data, CDC genomic data, and Facebook survey have been combined to inform agent interactions in epidemiological models \cite{aylett-bullockJuneOpensourceIndividualbased2021a, romero2021public}. More generally, recent works have combined differentiable ABMs with deep neural networks to effectively leverage heterogeneous data sources during the calibration process ~\cite{chopraDifferentiableAgentbasedEpidemiology2023b, quera2023bayesian}. 

Recent improvements in ABM scalability have been matched by an increased risk to individual privacy \cite{assefaGeneratingSyntheticData2021}, which has already manifested in the form of data leaks \cite{coxTMobilePutMy2019, IndonesiaProbesSuspected2021, kelleyIllinoisBoughtInvasive2021}. Statistical privacy methods such as K-anonymity and differential privacy have been employed \cite{bettini2009anonymity, wang2018privacy} to protect individual information on an aggregate level. Such methods have been used to release US census data \cite{bureauWhyCensusBureau} whilst protecting individual information at an aggregate level. Similar methods were also employed by Google and SafeGraph in the release of mobility trace data~\cite{aktay2020google}. Although statistical hiding methods are good at hiding identifiable information in aggregate statistics, they have limited capability of achieving privacy when applied at the level of individuals. This creates a difficulty in simulating population networks, as state propagation relies on both accumulating individual agents' states, whilst evaluating policy interventions relies on access to mobility data for each agent. As a result, statistical hiding methods provide poor privacy-utility trade-offs in the context of ABM. 

In contrast, cryptographic protocols have been used to provide strong privacy guarantees at the level of individuals. Recently, MPC has been explored for secure federated learning~\cite{mugunthanSMPAISecureMultiParty} and training of distributed graph neural networks~\cite{wangSecGNNPrivacyPreservingGraph2023}. These are constrained by high computation and communication cost of executing operations on deep neural networks, implementation complexity of secure aggregation, and intractability of secure backpropagation through non-linear mechanisms like graph attention. This requires hybrid approaches that leverage MPC with trusted execution environments for training DNNs \cite{jieMultiPartySecureComputation2022}. In contrast, the simplicity of mechanistic models, like ABMs, allows leveraging MPC in fully decentralized scenarios. RIPPLE\cite{guntherPrivacyPreservingEpidemiologicalModeling2022} introduces a Private Information Retrieval (PIR) based method to collect aggregate statistics on contact-tracing systems while preserving user privacy. \cite{frias2011agent} uses encrypted personal information, adding a layer of anonymity between the simulator and the agents. However, there is no mechanism for privacy-preserving calibration and interventions on ABMs \cite{assefaGeneratingSyntheticData2021}. To our knowledge, our work constitutes the first protocol that enables simulation, calibration, and analysis of ABMs while preserving agent privacy at every step.

%\textcolor{red}{good survey of MPC: https://eprint.iacr.org/2020/300.pdf - goes in 7.2}
%
%\bigskip
%
%\textcolor{red}{Some useful references for scale:\\ a) https://arxiv.org/pdf/2207.09714.pdf - mentioned above\\ b) https://web.media.mit.edu/~ayushc/motivation.pdf - could go in ABM prelim\\ c)https://web.media.mit.edu/~ayushc/AgentTorch.pdf - could go in future work, contains some ref for example realms for ABM use,\\ d) https://dl.acm.org/doi/10.5555/3545946.3598853 - mentioned above}
%
%\bigskip
%
%\textcolor{red}{add more refs and details.. Some relevant jp morgan papers: a) (https://dl.acm.org/doi/abs/10.1145/3383455.3422554 - goes in 7.2)  mentions there are no privacy-preserving calibration algorithms for ABMs. b) Private dark pools for secure trading without revealing trader's info (https://dl.acm.org/doi/abs/10.5555/3398761.3398969 - goes in 7.2) - Use these refs for data privacy section below.}
%

%% file: content/conclusions.tex
\section{Discussion and Conclusion}
In this paper, we introduced a paradigm that enables the simulation, calibration, and analysis of agent-based simulations on real-world data, while safeguarding the privacy of the involved agents. Our approach leverages MPC techniques to develop robust privacy-preserving protocols, without compromising the accuracy of the ABM output. We demonstrated the efficacy of our paradigm by presenting a case study in the city of Oxford, where we evaluated mask-wearing policies, performed gradient-assisted calibration to ground-truth data, and analyzed simulation outcomes using a privacy-preserving ABM.

The presented concept model can be further developed by extending the MPC protocols in multiple ways, including 
\begin{enumerate}
    \item Generalizing to higher-order networks to capture more complex contagion models. This could help understand the influence of group dynamics on social behavior. 
    \item Addressing practical engineering challenges, such as minimizing communication overheads, supporting asynchronous message passing, and leveraging distributed computing to accelerate computation.
    \item Combining MPC with federated learning. While MPC enabled simulating with individual agents, federated learning can help incorporate siloed institutional agents. For example, this would help privately integrate CDC genomic data or FED insurance claims into epidemiological and economics models.
    % This would allow access to siloed institutional data such as CDC genomic data or insurance claims, by extending the privacy guarantees to institutional agents, in a similar way that MPC does for individual agents. Under such privacy definition, federated learning can 
    
    % This would allow access to siloed institutional data such as CDC genomic data or insurance claims. While MPC unlocked individual data, we should generalize the calibration protocol to also incorporate siloed institutional data. For example, epidemiological ABMs can now use personal health information (with MPC) but, they can also benefit from using genomic data at CDC, insurance claims data at Census and direct deposit data at the Fed etc. This will require a new privacy definition to enable federated learning between institutional agents along with multi-party computation between individual agents.
\end{enumerate}
In the long term, private ABMs can be deployed through mobile devices, enabling passive and secure monitoring of actions and interactions within large populations. This would enable conducting digital experiments on the real-time behavior of complex systems with actual populations while ensuring robust security measures.